\documentclass[sigconf,screen]{acmart}

\setcopyright{cc}
\setcctype{by}
\acmDOI{10.1145/3832783.3834378}
\acmYear{2026}
\copyrightyear{2026}
\acmISBN{979-8-4007-2882-2/2026/10}
\acmConference[ASE '26]{Proceedings of the 41st IEEE/ACM International Conference on Automated Software Engineering}{October 12--16, 2026}{Munich, Germany}
\acmBooktitle{Proceedings of the 41st IEEE/ACM International Conference on Automated Software Engineering (ASE '26), October 12--16, 2026, Munich, Germany}
\acmSubmissionID{ase26main-p1085-p}
\received{2026-03-26}
\received[accepted]{2026-06-18}

\makeatletter
\def\ACM@cc@type{by}
\makeatother

\usepackage[utf8]{inputenc}
\usepackage[T1]{fontenc}
\usepackage{microtype}

\usepackage{enumitem}
\newlist{tightitemize}{itemize}{3} 
\setlist[tightitemize]{label=\labelitemi, noitemsep, leftmargin=*, topsep=0pt, partopsep=0pt}

\usepackage{xspace}

\usepackage[colorinlistoftodos,textwidth=15mm,textsize=tiny,obeyFinal]{todonotes}

\usepackage{xcolor}
\newcommand{\hldel}[1]{{%
    \setlength{\fboxsep}{1.5pt}%
    \colorbox{red!10}{\color{red!70!black}#1}%
}}
\newcommand{\hladd}[1]{{%
    \setlength{\fboxsep}{1.5pt}%
    \colorbox{green!10}{\color{green!50!black}#1}%
}}
\newcommand{\hlcmt}[1]{{%
    \setlength{\fboxsep}{1.5pt}%
    \colorbox{gray!10}{\color{gray!80}#1}%
}}

\usepackage[skip=4pt]{subcaption}
\usepackage{tabularray}

\NewTblrTheme{mockcaption}{
\SetTblrStyle{caption-tag}{font=\normalsize\bfseries}
\SetTblrStyle{caption-sep}{font=\normalsize\bfseries}
\SetTblrStyle{caption-text}{font=\normalsize\bfseries}
\SetTblrOuter{headsep=4pt}
}

\usepackage[frozencache,cachedir=_minted]{minted}
\newminted{diff}{
        numbersep=1pt, 
        autogobble, 
        fontsize=\scriptsize,
        breaklines=true,
        breakanywhere,     
        breakautoindent=true, 
        bgcolor=gray!10,
        bgcolorpadding=0.5em,
        escapeinside=@@, 
}

\newminted{python}{
    linenos,               
    numbersep=4pt,         
    autogobble,            
    fontsize=\scriptsize,
    breaklines=true,       
    breakanywhere,         
    breakautoindent=true,  
    escapeinside=!!,       
    python3=true,          
}

\usepackage[skins]{tcolorbox}
\newtcolorbox{findingbox}{
    enhanced,
    boxrule=0pt,
    frame hidden,
    borderline west={2pt}{0pt}{blue!60!black}, 
    colback=gray!10,                            
    sharp corners,                             
    fontupper=\itshape, 
    left=2pt, 
    right=0pt,
    top=1pt,
    bottom=1pt,
    beforeafter skip=1em,
}

\usepackage{stfloats}  

\makeatletter
\AtEndPreamble{%
  \global\@ACM@balancefalse
  \RequirePackage{pbalance}%
}
\makeatother

\begin{document}


\title{Understanding Bugs in Modern Agentic Frameworks: A Study of Symptoms, Root Causes, and Triggering Conditions}

\author{Xiaowen Zhang}
\orcid{0009-0009-4475-9442}
\affiliation{%
  \institution{Concordia University}
  \city{Montreal}
  \country{Canada}
}
\email{xiaowen.zhang@mail.concordia.ca}

\author{Hannuo Zhang}
\orcid{0009-0005-7211-2233}
\affiliation{%
  \institution{Concordia University}
  \city{Montreal}
  \country{Canada}
}
\email{hannuo.zhang@mail.concordia.ca}

\author{Shin Hwei Tan}
\authornote{Corresponding Author}
\orcid{0000-0001-8633-3372}
\affiliation{%
  \institution{Concordia University}
  \city{Montreal}
  \country{Canada}
}
\email{shinhwei.tan@concordia.ca}



\newcommand{\langchain}{LangChain\xspace}
\newcommand{\langgraph}{LangGraph\xspace}
\newcommand{\crewai}{CrewAI\xspace}
\newcommand{\autogen}{AutoGen\xspace}
\newcommand{\smolagents}{SmolAgents\xspace}

\newcommand{\shinhwei}[1]{\noindent\textcolor{red}{TODO: #1\xspace}}
\newcommand{\xiaowen}[1]{\nbc{XW}{#1}{blue}:}

\newcommand{\etal}{{et al.}\xspace}
\newcommand{\ie}{{i.e.},\xspace}
\newcommand{\eg}{{e.g.},\xspace}

\newcommand{\datacount}{409\xspace}

\begin{abstract}
Modern agentic frameworks such as \crewai and \autogen have evolved into complex, autonomous multi-agent systems, introducing reliability challenges that go beyond earlier pipeline-based LLM libraries. However, existing empirical studies focus on earlier LLM libraries or task-level bugs, leaving the unique complexities of these agentic frameworks unexplored.
We present a comprehensive study of \datacount fixed bugs across five representative agentic frameworks, proposing a five-layer architectural abstraction.
Our taxonomy identifies previously unreported symptom categories---Unexpected Execution Sequence, User Configuration Ignored, and Incomplete/Incorrect Trace---and isolates agent-specific root causes including Model-Related Fault, Cognitive Context Mismanagement, and Orchestration Fault. Notably, the model integration layer is the most bug-prone yet receives disproportionately low test inclusion rate during bug fixing (47\%), revealing a critical validation gap. Despite varying design paradigms, bug symptoms, root causes, and bug-prone components show substantial cross-framework consistency (JS similarity 0.62--0.88).
Finally, we present the first systematic study of bug-triggering conditions, identifying error-prone factor combinations across element configurations, input patterns, and operations, and demonstrate their transferability across frameworks, providing a foundation for test oracle design and cross-framework benchmark.
\end{abstract}

\maketitle

\section{Introduction}

With the recent advancement of large language models (LLM), several modern agent orchestration frameworks such as LangChain, LangGraph, CrewAI, AutoGen, and SmolAgents have emerged, each with distinct characteristics. These frameworks enable developers to compose autonomous agents, coordinate multi-agent workflows, manage contextual reasoning, and invoke external tools at scale. As they form the backbone of a growing ecosystem of downstream applications, defects within these frameworks can propagate widely, causing failures that range from silent semantic errors to system crashes.  
Therefore, it is important for developers and researchers to have a systematic understanding of the nature, causes, and triggering inputs of these defects.


A recent empirical study of defects in three LLM agent frameworks (i.e., LangChain, LlamaIndex, and Haystack)~\cite{xue2025llmbugstudy} proposed a taxonomy of nine root cause categories and six symptom categories, along with a four-component architectural abstraction: core schema, data preprocessing, agent construction, and featured modules. The study established a baseline by identifying Code Logic Issues (38\%) and API Misuse (24.7\%) as dominant root causes, Crash (31.8\%) and Incorrect Functionality (30.7\%) as the most prevalent symptoms. It further highlighted Unexpected Output (16.5\%) as a characteristic symptom, reflecting the probabilistic and generative nature of LLM-based systems. Its abstractions center on data pipelines, retrieval-augmented generation (RAG), and prompt engineering. Despite these contributions, the prior study has three key limitations that motivate our work. 
\textbf{First, it targets an earlier generation of LLM frameworks} whose abstractions center on
four components: i) data preprocessing, ii) core schema, iii) agent construction, and iv) featured modules --- a design philosophy that no longer reflects the current state of the ecosystem. Modern agentic architectures have shifted toward autonomous multi-agent execution, long-running stateful workflows, and direct model orchestration. Frameworks such as \autogen, \crewai, and \smolagents introduce qualitatively distinct paradigms such as event-driven messaging, role-based agent teams, and code-executing agents, that lead to qualitatively different failure modes~\cite{devstudyagentsystem}. Consequently, the previous four-component model, which is designed based on traditional software library structures, fails to capture the complexities of modern agent coordination, dynamic task delegation, and emergent runtime behaviors that characterize modern agentic systems. 

\textbf{Second, the study does not analyze bug-triggering scenarios}, a critical gap for advancing automated testing of agentic frameworks. Knowing not only \textit{what} bugs occur but \textit{under what conditions} they manifest is essential for constructing fault-revealing test inputs, designing test oracles, and supporting fault localization in multi-agent settings where non-deterministic interactions make reproduction inherently challenging.

\textbf{Third, the study does not examine the cross-framework transferability of bug characteristics}, leaving open whether the identified taxonomy and patterns generalize across frameworks with fundamentally different design philosophies. A transferability analysis is essential to distinguish framework-specific anomalies from universal bug patterns, and to inform the development of framework-agnostic repair and testing techniques. 

To address these gaps, we present an empirical study of \datacount fixed bugs from five representative frameworks: LangChain, LangGraph, CrewAI, AutoGen, and SmolAgents. To better reflect modern agentic architectures, we propose a refined five-layer abstraction comprising the Orchestration, Intelligence, Knowledge, Action, and Infrastructure layers. 
Our study aims to answer the following six Research Questions (RQs):

\begin{description}[leftmargin=*]
    \item[RQ1 (Symptoms):] \textit{What are the prevalent symptoms of bugs in modern agentic frameworks? } 
    Analyzing symptoms identifies new failure modes in modern agentic frameworks to aid the development of effective test oracles.

    \item[RQ2 (Root Causes):] \textit{What are the common root causes of these bugs?}
    Uncovering root causes reveals fundamental design flaws and challenges stemming from autonomous orchestration and LLM integration that compromise framework reliability.

    \item[RQ3 (Bug-prone Components):] \textit{Which architectural layers are most susceptible to defects, and which layers do developer testing efforts primarily focus on?}
    This analysis identifies bug-prone components and reveals potential misalignment between developer testing focus and the actual distribution of bugs.

    \item[RQ4 (Commonality \& Association):] 
    \textit{Do symptoms, root causes, and components exhibit commonality across frameworks, and what are the relationships between these dimensions?}
    This analysis reveals systemic failures 
and guides developers in prioritizing testing for critical failure patterns.

    \item[RQ5 (Triggering Scenarios):] \textit{What user-side scenarios are most likely to trigger bugs in agentic frameworks?}
    Understanding these scenarios helps guide the design of automated test generation tools that effectively target common failure-inducing inputs.

    \item[RQ6 (Transferability):] \textit{To what extent are bug-triggering scenarios transferable across different framework designs?}
    Exploring transferability facilitates the generalization of localized bugs as framework-agnostic benchmarks for systematic cross-platform validation.

\end{description}

Our analysis reveals a misalignment in agentic framework reliability, where the model integration layer exhibits the highest bug density (25\%) but comparatively low test inclusion rate during bug fixing (47\%). Furthermore, despite varying design paradigms across the selected frameworks, their bug symptoms, root causes, and bug-prone components show high distributional similarity (JS similarity 0.62–0.88), revealing a shared reliability landscape.



Specifically, this paper makes the following contributions:

\begin{description}[leftmargin=*]
    \item[Architectural Abstraction for Modern Agentic Frameworks.] We propose a five-layer architectural abstraction grounded in a systematic analysis of \datacount fixed bugs across five representative modern agentic frameworks, providing a structured foundation for component-level testing and cross-framework comparison.


    \item[Taxonomies of Agentic Failures.] We develop a taxonomy to capture failures characteristic of modern agentic frameworks, identifying new symptoms such as Unexpected Execution Sequence, User Configuration Ignored, and Incomplete/Incorrect Trace, and isolating agent-specific root causes including Model-Related Fault, Cognitive Context Mismanagement, and Orchestration Fault.
    

    \item[Bug-Triggering Conditions and Transferability.] We present \\ 
    the first systematic study of bug-triggering conditions in agentic frameworks, identifying error-prone factors across element configurations, input patterns, and operations. We further explore the cross-framework transferability of these conditions, providing actionable insights for proactive defense and robust benchmarking.
    
\end{description}
\section{Background and Related Work}
\label{sec:bg}

\subsection{Evolution of Agentic Frameworks}
\label{sec:bg-evol}
LLM application development has transitioned from using simple utility libraries to relying on complex autonomous orchestration infrastructures~\cite{devstudyagentsystem}. Early frameworks~\cite{yao2023react, Liu_LlamaIndex_2022, haystack,langchain} served as stateless intermediaries, providing standardized interfaces for prompt construction, retrieval-augmented generation pipelines, and basic API wrapping.  In this paradigm, frameworks served as modular tool collections rather than cohesive execution environments, leaving developers to manually manage conversation state and multi-step workflows. In contrast, modern agentic frameworks such as \langgraph~\cite{langgraph}, \crewai~\cite{crewai}, \autogen~\cite{wu2024autogen}, and \smolagents~\cite{smolagents} have introduced integrated runtime environments characterized by persistent state management and autonomous decision loops. Unlike earlier approaches, they support cyclic control flows and coordination, realized through patterns such as event-driven messaging, role-based agent teams, and sandboxed code execution. 
This evolution shifts the role of frameworks from a passive library to an active orchestrator managing context persistence and dynamic tool invocation. Consequently, system correctness becomes tightly coupled to the framework's coordination logic and state transition semantics.

\subsection{Study Scope and Framework Selection}

To ensure a rigorous empirical analysis, we classify bugs into three levels: framework-level, application-level bugs, and task-level failures. We define \textbf{framework-level bugs} as defects within the source code of agentic frameworks, such as recovery errors, that occur independently of model capabilities. In contrast, \textbf{application-level bugs} stem from the misuse of framework APIs or errors in user-defined logic during system construction. Distinct from software defects, \textbf{task-level failures} refer to behavioral issues (\eg hallucinations) where the agent fails an objective despite the code operating as intended, often representing the inherent uncertainty of model outputs. This three-tier distinction is essential for isolating the orchestration infrastructure from both application logic and model performance.

We selected five agentic frameworks from a recent survey \cite{compareframeworks} to represent diverse architectures and abstraction levels. 
\textbf{\langchain} typifies linear and directed acyclic graph (DAG) orchestration, while \textbf{\langgraph} introduces stateful cyclic graphs for iterative reasoning.
To represent multi-agent collaboration, we include \crewai, which prioritizes structured process-oriented workflows, and \autogen, which facilitates autonomous conversation-driven interactions. 
Finally, \textbf{\smolagents} represents a code-centric paradigm where agents interact primarily through sandboxed code executors.



\subsection{Modern Agentic Framework Abstraction}

\begin{figure}[t]
    \centering
    \includegraphics[width=0.4\textwidth]{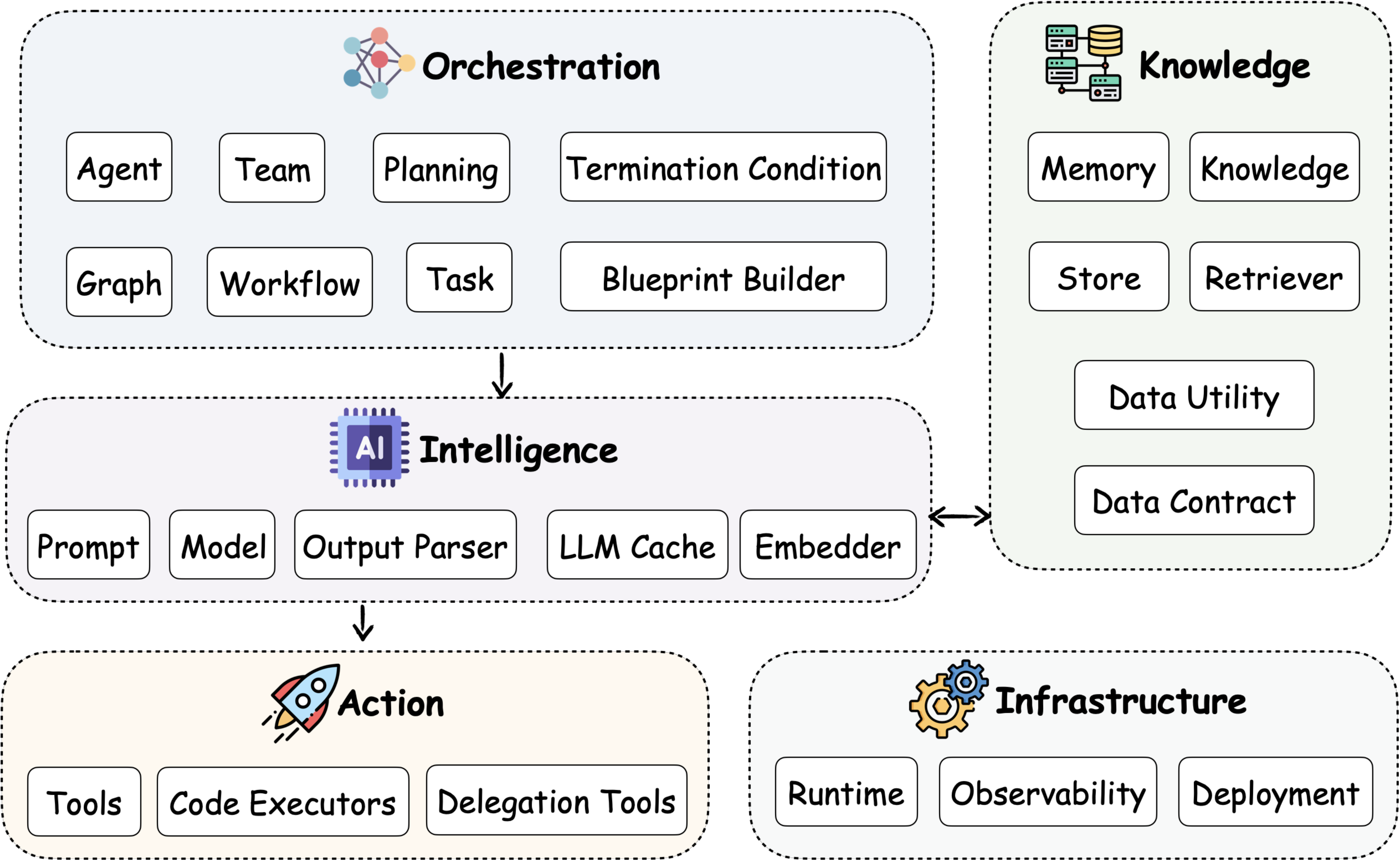}
    \caption{Conceptual architecture of agentic frameworks.}
    \label{fig:architecture}
\end{figure}

To systematically analyze diverse agentic frameworks, we propose a five-layer architectural model (Figure~\ref{fig:architecture}), spanning from low-level environment management to high-level multi-agent coordination.
We derived this abstraction by synthesizing conceptual components from recent literature~\cite{derouiche2025agentic} and functional modules in the five studied frameworks.
Unlike the prior study that groups framework-specific modules into a flat ``Featured Modules'' category~\cite{xue2025llmbugstudy}, our abstraction redistributes them into functional layers, capturing the integrated nature of agentic systems. 

\noindent\textbf{Infrastructure} provides the runtime backbone for execution loops, inter-agent communication, lifecycle management, and monitoring, ensuring reliable operation and visibility into agent workflows. 

\noindent\textbf{Action} interfaces agents with the external environment via tools, code executors, and delegation mechanisms that translate agent decisions into concrete operations. 

\noindent\textbf{Knowledge} manages context, state, and data through memory stores, retrievers, and data contracts, supporting state persistence and consistent information flow across agents and workflows.

\noindent \textbf{Intelligence} serves as the cognitive engine, managing inference, semantic mapping, and protocol adaptation through models, adapters, prompts, embedders, and output parsers that convert raw LLM responses into structured, actionable representations. 

\noindent\textbf{Orchestration} defines agents, task structures, and collaboration topologies, governing dynamic execution through workflow and graph-based connections, planning, termination conditions, and team management, with construction tooling to instantiate these structures from configurations or blueprints.

\section{Methodology}
\label{sec:method}

\subsection{Data Collection and Annotation}

We collected bug-related issues with confirmed fixes from the official GitHub repositories of the selected agentic frameworks using a systematic pipeline to identify traceable issue–fix pairs. We mined all closed issues labeled as bugs, sorted by creation time from newest to oldest. Then, we filtered out issues without a ``completed'' \texttt{state\_reason}, excluding those unlikely to have fixes, such as not-planned issues. To identify corresponding fixes, we retrieved timeline events for each issue and collected cross-referenced pull request (PR) links. Due to GitHub API rate limits, we examined only the first 100 events per issue and retained the latest 500 issues with linked PRs per framework. By our collection cutoff date (August 11, 2025), this process yielded 820 candidate issues: \langchain 419, \langgraph 25, \crewai 96, \autogen 183, and \smolagents 97. Given the large number of issues in \langchain, we randomly sampled 100 \langchain issues to prevent one framework from dominating the dataset while preserving internal representativeness, and combined them with all others, resulting in 501 issues for manual verification.
During manual verification, we kept only issues with PRs that fixed bugs in the core framework implementation, excluding duplicates, multi-bug reports, issues without fixes, or fixes only affecting peripheral tools (\eg \crewai CLI and \autogen Studio) or test code. Multi-bug reports were excluded because a single report may contain multiple independent bugs, complicating one-to-one bug annotation. After filtering, 409 issues remained:  \langchain 88, \langgraph 17, \crewai 77, \autogen 146, and \smolagents 81. 
As \langgraph is built on \langchain components and they are often used together, 
we collectively refer to both frameworks as LC/LG. 

We adopt precise definitions for two key terms~\cite{ieeestd1990,swebok2024,li2017tos}: a \emph{fault} is a static defect in code or configuration, and a \emph{failure} is an externally observable incorrect behavior. These terms correspond to our root cause and symptom dimensions, respectively. Other terms (\emph{bug, defect, error, issue}) are used in their general sense.

For each bug report, we systematically annotated four dimensions: symptoms, root causes, buggy components, and triggering patterns. Following prior work~\cite{dlframework2023,huaien2024,haibo2025}, we derived our taxonomy using an open-coding approach. One author first reviewed all issues and associated fixes to establish an initial set of categories. 
Two authors independently annotated the collected issues across 11 batches, covering 5\% of the data in the first two batches and 10\% in each subsequent iteration.
After each batch, we measured inter-annotator agreement using Cohen's kappa ($\kappa$)~\cite{cohenkappa2010} and resolved disagreements through discussion, iteratively refining the categories and involving a third author when consensus could not be reached.
The average $\kappa$ was 0.68 in the first iteration, improved to 0.96 in the second, and consistently exceeded 0.89 in all remaining iterations.
This manual analysis spanned six person-months as it required extensive documentation review to gain a deeper understanding of agentic frameworks.

\subsection{Analysis Methods}
To address RQ1, RQ2, and RQ3, our study developed taxonomies for bug symptoms, root causes, and components by analyzing the \datacount bug reports and their associated fixes. For RQ3, we determined the affected conceptual components based on the functionality of the modified code and checked if each bug fix had corresponding test cases. 
Then, we further analyzed the commonality and association among these dimensions (RQ4).

\noindent \textbf{Commonality analysis:} To assess cross-framework consistency of bug characteristics, we evaluated the distributional similarity of symptoms, root causes, and components across the selected frameworks. While prior studies~\cite{shen2021dlcompiler, liu2026llminferengine, dlframework2023} used Spearman or Pearson correlation, these metrics can be misleading. 
Specifically, rank-based Spearman is sensitive to minor permutations in low-frequency categories, while linear Pearson is skewed by dominant categories. To obtain a more robust measure, we employed Jensen-Shannon Divergence (JSD)~\cite{jensenshannon}, which measures the overlap between two probability distributions by comparing each to their mean. We then define \textit{Jensen-Shannon (JS) similarity} as $1 - \sqrt{JSD}$, ranging from 0 (completely different) to 1 (identical).

\noindent \textbf{Association analysis:} To investigate relationships among symptoms, root causes, and components, we followed the statistical approach in \cite{haibo2025}. First, we applied the Chi-squared test of independence to identify significant associations between dimension pairs (\eg symptoms and root causes). We then used Cramér's $V$ to measure association strength. Following Cohen's guidelines~\cite{cohen2013statistical}, we interpreted $V$ values as weak (0.1--0.2), moderate (0.2--0.3), or strong (>0.3).
To identify significant category pairs, we conducted a post-hoc analysis with a Bonferroni correction~\cite{shan2017fisher}, ensuring only associations unlikely by chance were marked significant. 

To answer RQ5, we analyzed bug-triggering scenarios for 273 bugs with reproduction code or descriptive configuration. 
Unlike a prior study on refactoring engines~\cite{haibo2025} where bugs often arise from simple input code, agentic frameworks exhibit complex interactions between element configurations, task types, and runtime behaviors. We characterized these via three sub-dimensions: Element Configurations (251 bugs), Input Patterns (56), and Operations (101). We define a \emph{factor} as a bug-triggering characteristic, where a single bug may involve multiple factors per sub-dimension.
Specifically, \textbf{Element Configurations} involve static element settings and environment variables, typically represented as element-attribute pairs (\eg \texttt{model:modelId}). \textbf{Input Patterns} characterize runtime data such as task semantics or LLM responses.  \textbf{Operations} denote functional behaviors, such as explicit tool calls, that exceed standard routines (\ie basic inference passes).

\noindent \textbf{Automated Frequent Pattern Mining:} To identify recurring bug-triggering patterns, we applied FP-Growth~\cite{han2004fpgrowth}, a technique proven effective for frequent pattern mining in log analysis~\cite{,fpgrowthinlog,fpgrowthfailure}. 
Each bug is treated as a transaction of factors, and an FP-tree is constructed to compactly encode shared structure. Mining this tree reveals frequent co-occurring factor sets above a support threshold, allowing us to characterize common multi-factor bug-triggering conditions.
For factor combinations, we adopted a mixed-granularity approach to capture patterns across hierarchical levels. For any factor in the form $a:b$, we augmented the set by injecting its parent factor $a$. 
We set FP-Growth \texttt{min\_support} to 0.05, requiring a pattern to appear in at least 5\% of cases, and excluded patterns with fewer than five occurrences.

\noindent\textbf{Transferability:}  To answer RQ6, we investigated whether bugs from one agentic framework recur in others. Given their shared analogous abstractions and coordination logic, we hypothesized cross-framework manifestations and sampled 35 representative bugs from the three triggering dimensions.
Specifically, we randomly sampled bugs for the Element Configurations, 
while for Input Characteristics and Operations, we prioritized bugs whose triggering patterns are defined by \textit{shared agentic interaction surfaces} (\ie common abstractions such as models and agents, including their configurations and operations)---specifically message triggers and serialization interfaces---rather than framework-specific internals, as these provide structural preconditions for cross-framework reproduction.
We evaluated these scenarios against the latest versions of \langchain (1.2.7), \langgraph (1.0.7), \crewai (1.9.2), \autogen (0.7.5), \smolagents (1.24.0). 
For each source bug, we evaluated its presence across the four remaining frameworks. For each target framework, we first searched its GitHub repository for existing reports. A target issue was categorized as \textit{Duplicate} if its reproducible code was functionally equivalent to the source.  If no duplicate existed, we manually adapted the source bug code to the target framework. Upon successfully triggering the bug, we submitted a report and recorded it as an \textit{Adapted} issue. If adaptation failed, we searched for a \textit{Variant} issue sharing the same failure pattern based on the original bug-triggering factors but differing in component or context. For each source bug, only the highest-priority relation was recorded per target framework. A source bug is considered \emph{transferable} if at least one target framework contains a Duplicate, Adapted, or Variant issue.
{
\section{RQ1: Symptoms}
\label{sec:symptom}


\subsection{Symptom Categories}
Our analysis identified nine distinct symptom categories below:

\noindent \textbf{Crash/Error.} The framework or its internal components throw an exception, as indicated by error messages or stack traces. 

\noindent \textbf{Initialization Failure.}
Failures preventing framework initialization, involving module import, installation, or syntax errors. 

\noindent \textbf{Unexpected Intermediate State.} A transient internal state produced during execution deviates from expectations. Users typically observe such deviations in logs, debug outputs, or code, and report them without mentioning any resulting consequences. 
}

\noindent \textbf{Unexpected Output.} 
There is a mismatch between the expected and actual results,
including responses from LLMs, agent decisions, workflow outcomes, tool and utility results, or serialized exports. 
For example, \smolagents-1481 illustrates an unexpected agent decision, where the manager prematurely returned an incomplete answer before collecting all parallel subtask results~\cite{smol1481multicall}.

\noindent \textbf{Unexpected Execution Sequence.}  The framework produces an interaction sequence that deviates from the intended agent interaction order or task progression, such as order mismatch, early exit, unexpected continuation, or missing tool calls. For instance, in \autogen-6710, the workflow manager mishandled loops, preventing any agents from activating and causing an early exit~\cite{ag6710multicycles}.

\noindent \textbf{Incomplete/Incorrect Trace.} The framework generates partial or misformatted execution traces or logs, hindering analysis and debugging. A case is \autogen-6531, where the log of an \texttt{LLMCallEvent} omitted available tools, losing crucial execution context~\cite{ag6531trace}.   

\noindent \textbf{Performance Anomaly.} The framework exhibits degraded performance or unresponsiveness, such as slow execution or hangs.

\noindent \textbf{User Configuration Ignored.} The framework silently ignores user-provided  configurations, rendering them ineffective. For example, in \langchain-32059~\cite{lc32059datapropagation}, 
setting \texttt{num\_gpu=4} for \texttt{OllamaEmbed\-dings} had no effect, with all computations falling back to CPU.

\noindent \textbf{Others.} This includes rare or miscellaneous symptoms that do not fit into other categories (e.g., broken documentation links).

\subsection{Distribution and Comparison with Baseline}
\label{sec:symptom-analysis}

\begin{table}[t]
    \centering
    \footnotesize
    \begin{talltblr}[
        theme = mockcaption,
        caption = {Bug distribution by symptoms and frameworks.},
        label = {tab:symptom},
        remark{Abbreviations} = {
            \textit{Smol} denotes SmolAgents.
            \textit{Unexp. State} and \textit{Unexp. Sequence} denote Unexpected Intermediate State and Unexpected Execution Sequence, respectively.
        },
    ]{
            width = \columnwidth,
            colspec = {l r r r r r r r r r r}, 
            cells = {m},
            colsep = 1.5pt, 
            rowsep = 1.5pt, 
            cell{1}{2,4,6,8,10} = {c=2}{m}, 
            cell{1}{1} = {r=2}{m},
            hline{1,2,3,Z} = {solid}, 
            vline{2,6,8,10} = {solid}, 
            vline{4} = {0.8pt, solid},
            row{even} = {bg=gray!10}, 
            row{1,2} = {c, font=\bfseries, bg=gray!30},
            cell{3,4,5}{2,3} = {font=\bfseries},
            cell{3,4,5}{4,5} = {font=\bfseries},
            cell{3,4,6}{6,7} = {font=\bfseries},
            cell{3,4,6}{8,9} = {font=\bfseries},
            cell{3,4,5}{10,11} = {font=\bfseries},
        }
        Symptom  & Total & & LC/LG & & \crewai & & \autogen & & Smol & \\
        & \# & \% & \# & \% & \# & \% & \# & \% & \# & \% \\
        Crash/Error & 222 & 54.28 & 54 & 13.20 & 34 & 8.31 & 78 & 19.07 & 56 & 13.69 \\
        Initialization Failure & 47 & 11.49 & 7 & 1.71 & 17 & 4.16 & 16 & 3.91 & 7 & 1.71 \\
        Unexpected Output & 41 & 10.02 & 21 & 5.13 & 3 & 0.73 & 10 & 2.44 & 7 & 1.71 \\
        Unexp. State & 29 & 7.09 & 5 & 1.22 & 6 & 1.47 & 16 & 3.91 & 2 & 0.49 \\
        Unexp. Sequence & 20 & 4.89 & 4 & 0.98 & 5 & 1.22 & 8 & 1.96 & 3 & 0.73 \\
        User Configuration Ignored & 14 & 3.42 & 5 & 1.22 & 4 & 0.98 & 3 & 0.73 & 2 & 0.49 \\
        Incomplete/Incorrect Trace & 12 & 2.93 & 2 & 0.49 & 3 & 0.73 & 5 & 1.22 & 2 & 0.49 \\
        Performance Anomaly & 9 & 2.20 & 3 & 0.73 & 2 & 0.49 & 3 & 0.73 & 1 & 0.24 \\
        Others & 15 & 3.67 & 4 & 0.98 & 3 & 0.73 & 7 & 1.71 & 1 & 0.24 \\
    \end{talltblr}
\end{table}

Table~\ref{tab:symptom} presents the frequency and percentage of bug symptoms across studied frameworks. 
\textbf{Crash/Error} (54\%) is the most common symptom, reflecting persistent challenges in execution stability, consistent with prior studies on framework bugs~\cite{dlframework2023,xue2025llmbugstudy}. Automated techniques like fuzzing remain effective for detecting these runtime failures.
\textbf{Initialization Failure} (11\%) ranks second, with most cases (30/47) arising from module import errors caused by missing dependencies, version conflicts, or incomplete framework module exports. Due to Python's interpreted nature and dynamic imports, these issues appear only at runtime, revealing gaps between configuration correctness and execution-time validation.
\textbf{Unexpected Output} (10\%) is the third most common symptom, particularly in LC/LG (5\%). Most cases (18/41) originate from core components, including model, agent, or workflow outputs. Such deviations may stem from the non-deterministic behavior of underlying models or orchestration defects within agents and workflows, making output validation particularly challenging as correctness often cannot be determined by simple assertions.


Compared to the baseline study~\cite{xue2025llmbugstudy}, we identify four isolated symptoms in modern agentic frameworks. 
Import errors, which the baseline study primarily links to crashes as root causes, are treated as symptoms of \textbf{Initialization Failure}. 
These account for 7.3\% of all bugs in our dataset, compared to 3.5\% in the baseline study, indicating a higher prevalence of dependency and environment setup issues in agentic frameworks.
\textbf{Unexpected Execution Sequence}, \textbf{User Configuration Ignored}, and \textbf{Incomplete/Incorrect Trace} are newly identified in modern frameworks, capturing failures in multi-agent orchestration, configuration enforcement, and trace recording (concerns that were absent or conflated with application-layer issues in earlier frameworks). These symptoms are treated as distinct categories (rather than refinements of the broader incorrect-functionality category as in~\cite{xue2025llmbugstudy}) because they capture failures in the execution process rather than the final output, representing concerns that were absent in earlier framework studies.
These findings reveal gaps in validating execution paths, configuration effects, and trace integrity in modern agentic systems.

\begin{findingbox}
\textbf{Finding 1:}
Crash/Error (54\%), Initialization Failure (11\%), and Unexpected Output (10\%) are the most prevalent symptom categories. Unexpected Execution Sequence, User Configuration Ignored, and Incomplete/Incorrect Trace are newly identified categories not reported in the baseline study.
\end{findingbox}
\section{RQ2: Root Causes}
\label{sec:root_cause}


Our analysis grouped the root causes of \datacount bugs into five categories: Agent-Specific Fault, General Programming Fault, Dependency/Environment Fault, Documentation Fault, and User Misuse.

\begin{figure*}[t]
    \begin{subfigure}[t]{0.49\textwidth}
            \begin{diffcode}
 def _process_create_args(self, messages): ...
+ @\hlcmt{\# Gemini only accepts a single system message}@
+  if create_args.get("model", "").startswith(@\hladd{"gemini-"}@):
+    messages = create_args["messages"] # list of messages
+    messages = @\hladd{self.\_merge\_system\_messages}@(create\_args["messages"])
   ...
            \end{diffcode}
            \caption{Model Request Incompatibility: \autogen-6116~\cite{ag6116multisysprompt} \label{fig:reqincomp}}
    \end{subfigure}
    \hfill 
    \begin{subfigure}[t]{0.49\textwidth}
            \begin{diffcode}    
 @\hlcmt{\# chunk1: \{"name":"search", "args":'\{"query":', "id":"1", "index":1\}}@
 @\hlcmt{\# chunk2: \{"name":None, "args":' "test"\}', "id":None, "index":0\}}@ 
 @\hlcmt{\# [Fix] Handle chunks where model omits name/id}@
+  elif (e.get("type") == "tool_call_chunk" and merged
+    and @\hladd{e.get("name") is None and e.get("id") is None):}@
+    merge\_special\_chunk(e, merged)
            \end{diffcode}
            \caption{Model Response Incompatibility: \langchain-31511~\cite{lc31511unexpoutput} \label{fig:respincomp}}
    \end{subfigure}
    \begin{minipage}{0.4\textwidth}
    \begin{subfigure}[t]{\textwidth}
        \begin{diffcode}
 def tool(tool_function: Callable) -> Tool:
   class SimpleTool(Tool):
-    @\hldel{def \_\_init\_\_(self, name:str, ..., function: Callable):}@
-      self.name = name
-      ...
-      self.forward = function
+    @\hladd{def \_\_init\_\_(self):}@
       self.is_initialized = True
       
   ...        
-  simple_tool=SimpleTool(name=tool_json_schema["name"], ..., function=tool_function)
-  ...
+  SimpleTool.name = tool_json_schema["name"]
+  ...
+  SimpleTool.forward = staticmethod(tool_function)
+  simple_tool = SimpleTool()
   return simple_tool
        \end{diffcode}
        \caption{Class Design Defect: \smolagents-913~\cite{smol913classdesign} \label{fig:classdesign}}
    \end{subfigure}%
    \end{minipage}
    \hfill%
    \begin{minipage}{0.59\textwidth}
        \begin{subfigure}[t]{\textwidth} 
        \begin{diffcode}
@\hlcmt{\# prompts/toolcalling\_agent.yaml}@
 system_prompt: |-
-  You can give @\hldel{requests}@ to team members...The only argument you can give is @\hldel{'request'}@.
+  You can give @\hladd{tasks}@ to team members...The only argument you can give is @\hladd{'task'}@.
        \end{diffcode}
        \caption{Prompt Fault: \smolagents-606~\cite{smol606prompterr} \label{fig:promptfault}}
    \end{subfigure}
    \hfill
        \begin{subfigure}[t]{\textwidth}
            \begin{diffcode}
 runtime = Runtime(context=context, store=store, stream_writer=stream_writer, previous=None)
+parent_runtime = config[CONF].get(CONFIG_KEY_RUNTIME, DEFAULT_RUNTIME)
+runtime = @\hladd{parent\_runtime.merge(runtime)}@
            \end{diffcode}
            \caption{Context Propagation Fault: \langchain-5700~\cite{lg5700contextpropagate} \label{fig:contextpropagate}}
        \end{subfigure}
        \hfill
    \begin{subfigure}[t]{\textwidth}
            \begin{diffcode}
   def to_messages(self, summary_mode: bool, **kwargs) -> List[Message]: ...
     return [Message(role=MessageRole.ASSISTANT,content=[{"type":"text","text":self.plan}]),
+      Message(@\hladd{role=MessageRole.USER,}@content=[{"type":"text","text":"Execute the plan."}]),
     ]
            \end{diffcode}
            \caption{Message Fault: \smolagents-1097~\cite{smol1097contextmsg} \label{fig:contextmsg}}
        \end{subfigure}
    \end{minipage}
    \caption{Simplified code snippets illustrating representative root causes.}
\end{figure*}

\subsection{Agent-Specific Fault}
These faults reflect incorrect design decisions, assumptions, or execution policies in agentic frameworks, potentially disrupting agents' reasoning, perception, or orchestration. Fixing them requires domain knowledge, including interaction specifications and context propagation principles. We divide these faults into three categories: Model-Related Fault, Cognitive Context Mismanagement, and Orchestration Fault.

\noindent \textbf{Model-Related Fault.} These faults arise from misalignments between framework logic and LLM service chain, including the model, its API endpoint, and its client SDK (\eg LiteLLM~\cite{litellm}): 

\begin{tightitemize}
\item \textbf{Model Request Incompatibility (25/48):} 
The framework constructs inputs violating LLM-specific constraints (e.g., unsupported parameters or message formats), causing invalid request errors. 
In \autogen-6116 (Figure~\ref{fig:reqincomp}), Gemini-2.0-Flash 
responded only to the final system message when multiple were provided. This is fixed by merging all system messages.


\item \textbf{Model Response Incompatibility (19/48):} 
The framework fails to parse or merge model responses due to nondeterministic behavior or uncommon response schema. 
In \langchain-31511 (Figure~\ref{fig:respincomp}), the framework failed to merge tool call chunks from Qwen3 when subsequent chunks lacked \texttt{name} and \texttt{id}. The fix added merging logic to handle such omissions. 

\item \textbf{Other causes (4/48):} 
The framework fails due to incorrect model metadata, or unsupported features.
\end{tightitemize}

\noindent \textbf{Cognitive Context Mismanagement.} We define \emph{cognitive context} as runtime information shaping agents' situational awareness (e.g., conversation history, knowledge, tasks, tools, and collaborators). Mismanaging this context disrupts agent cognition, leading to degraded understanding, inconsistent decisions, or incorrect actions.
\begin{tightitemize}
\item \textbf{Prompt Fault (8/24):} Predefined prompts may contain faults (e.g., incorrect task specifications, insufficient instructions), leading to unexpected outputs. 
In \smolagents-606 (Figure~\ref{fig:promptfault}), a manager agent failed to delegate tasks because its system prompt had the wrong parameter (\texttt{request} instead of \texttt{task}) .

\item \textbf{Message Fault (6/24):} 
Errors in managing conversation history (e.g., missing messages, wrong roles,  history truncation). 
For example, in SmolAgents-1097 (Figure~\ref{fig:contextmsg}), exporting a planning step as an assistant message caused some LLMs to continue planning rather than execute tasks, as the missing role-change signal disrupted the agent's transition between task phases.

\item \textbf{Context Propagation Fault (10/24):} The framework fails to propagate/isolate cognitive context, resulting in inconsistent or stale context across steps, agents, or components.
For instance, \langchain-5700 (Figure~\ref{fig:contextpropagate}) reported that a subgraph could not access the runtime context of its parent graph, because the framework did not merge the parent context into the subgraph.

\end{tightitemize}

\noindent\textbf{Orchestration Fault.} These faults reflect incorrect coordination of 
agentic workflows, spanning navigation (workflow routing, speaker 
selection), termination (signal and condition handling), and 
component coordination (concurrency, mode transitions, action alignment). 
For instance, \autogen-6710~\cite{ag6710multicycles} exhibited premature 
termination due to incorrect cycle tracking, leaving prerequisites unsatisfied 
and blocking agent execution.


\subsection{General Programming Fault}
This encompasses common coding mistakes, divided into 12 sub-types. Following prior work~\cite{androidfuncbugs2023,exclusivecat2006}, we organize these according to descending scope, ranging from broad feature-level faults to narrow statement-level errors.


\noindent \textbf{Missing Case Handling.}
The framework does not implement specific features or edge cases. 
For example, in \autogen-5518, the framework did not support PowerShell code, producing an incorrect file extension, causing the local executor to reject the code~\cite{ag5518mscenario}.

\noindent \textbf{Class Design Defect.} 
Framework classes may violate architectural constraints (e.g., incorrect inheritance or invalid protocols). In \smolagents-913, \texttt{SimpleTool} broke validation rules by defining \texttt{\_\_init\_\_} parameters without defaults, causing errors when exported for sandbox execution. The fix (Figure~\ref{fig:classdesign}) removed these parameters and assigned them externally to ensure compliance.

\noindent \textbf{State/Resource Mismanagement.} This involves mishandled state updates or resource lifecycles, causing inconsistencies or failed operations. In \autogen-6308~\cite{ag6308lifecycle}, a \texttt{SearchClient} was closed prematurely by an \texttt{async with} block, causing later tool calls to fail.


\noindent \textbf{Concurrency Fault.} The framework mishandles concurrent execution (e.g., asynchronous tasks, thread-unsafe scheduling, or race conditions). In \autogen-4490~\cite{ag4490concurrency},
unawaited \texttt{WriteStateAsync()} caused out-of-order writes and \texttt{InconsistentStateException}.

\noindent \textbf{Serialization Fault.} The framework mishandles (de)serialization, violating high-level data contract assumptions (e.g., schema consistency, completeness, and idempotence). 

\noindent \textbf{API Misuse.} 
The framework violates API contracts by passing invalid or missing parameters, leading to invocation failures or ineffective calls.


\noindent \textbf{Missing Parameter Forwarding.} These faults occur when optional parameters fail to reach their targets due to omission or absence in the receiving interface.
Such faults rarely trigger explicit error messages but may silently affect downstream processing.


\noindent \textbf{Wrong Control Flow.} 
The framework executes in incorrect order or under unintended conditions due to incorrect control flow (e.g., evaluating the wrong branch~\cite{ag6730wcfcondition}). 

\noindent \textbf{Missing Check.} The framework executes existing logic without validating preconditions (\eg nullity, state, or existence checks), allowing invalid inputs or states to propagate.

\noindent \textbf{Incorrect Variable/Value.} 
The framework uses incorrect variables or values, mismanages local variable updates, or constructs values with incorrect structures.

\noindent \textbf{Type Fault.} The framework incorrectly uses or converts data types, resulting in type mismatches, invalid casting, or improper type conversions that violate expected type constraints.

\subsection{Dependency/Environment Fault} 
This category captures problems related to dependency management and execution environment compatibility, which affect framework integration, deployment, or extensibility.

\noindent \textbf{Dependency Fault.} 
These faults stem from missing or incompatible dependencies, third-party library bugs, or structural issues (\eg, improper module exports and rigid coupling) that hinder integration and extensibility.

\noindent \textbf{Environment Incompatibility.} These occur when the framework is incompatible with the execution environment (\eg, operating system, Python version, container image, hardware, or drivers).

\subsection{Documentation Fault}

This arises when documentation or tutorials misrepresent 
framework interfaces, behavior, or usage, encompassing incorrect examples 
(32/51), missing constraints (10/51), and outdated descriptions (9/51). 
For instance, \crewai-2647~\cite{crew2647docs} reported an authentication 
error when enabling planning for a \texttt{Crew} using non-OpenAI models, 
as the framework silently invoked \texttt{gpt-4o-mini} for planning without documenting the implicit OpenAI API key requirement.


\subsection{User Misuse}
This category includes issues caused by incorrect interactions or misconfigurations by users rather than internal framework faults. 

\subsection{Distribution and Comparison with Baseline}
\label{sec:rc-analysis}
\begin{table*}[!t]
    \centering
    \footnotesize
    \caption{The number of bugs due to different root cause in each agentic framework.}
    \label{tab:root-cause}
    \begin{tblr}{
        width = \textwidth,
        colspec = {l l r r r r r r r r r r},
        cells = {m}, 
        colsep = 3pt, 
        rowsep = 1.5pt, 
        cell{1}{1,2} = {r=2}{m},
        cell{1}{3,5,7,9,11} = {c=2}{halign=c}, 
        cell{3}{1} = {r=12}{m}, 
        cell{15}{1} = {r=4}{m},  
        cell{20}{1} = {r=3}{m},  
        cell{19}{1} = {c=2}{m},
        cell{Z}{1} = {c=2}{m},
        hline{1,2,3, 14, 15, 18, 19, 20, 22, 23, 24, 25} = {solid},
        vline{2,3,5,7,9,11} = {solid}, 
        vline{5} = {0.8pt, solid},
        row{even} = {bg=gray!10}, 
        row{22, 23} = {bg=white},
        column{1} = {bg=white},   
        row{14, 18, 22} = {font=\bfseries},
        row{1,2} = {font=\bfseries, bg=gray!30}, 
        cell{3,15,19}{3,4} = {font=\bfseries},
    }
        Root Cause & Subcategory & Total & & LC/LG &  & \crewai &  & \autogen &  & Smol &  \\
        & & \# & \% & \# & \% & \# & \% & \# & \% & \# & \% \\

        General Programming Fault & Missing Case Handling & 51 & 12.47 & 17 & 4.16 & 9 & 2.20 & 7 & 1.71 & 18 & 4.40 \\
         & Incorrect Variable/Value & 26 & 6.36 & 12 & 2.93 & 6 & 1.47 & 3 & 0.73 & 5 & 1.22 \\
         & Wrong Control Flow & 23 & 5.62 & 8 & 1.96 & 6 & 1.47 & 7 & 1.71 & 2 & 0.49 \\
         & Class Design Defect & 19 & 4.65 & 6 & 1.47 & 2 & 0.49 & 6 & 1.47 & 5 & 1.22 \\
         & Type Fault & 18 & 4.40 & 7 & 1.71 & 5 & 1.22 & 4 & 0.98 & 2 & 0.49 \\
         & State/Resource Mismanagement & 17 & 4.16 & 5 & 1.22 & 1 & 0.24 & 10 & 2.44 & 1 & 0.24 \\
         & Missing Check & 17 & 4.16 & 9 & 2.20 & 1 & 0.24 & 6 & 1.47 & 1 & 0.24 \\
         & API Misuse & 16 & 3.91 & 2 & 0.49 & 4 & 0.98 & 8 & 1.96 & 2 & 0.49 \\
         & Missing Parameter Forwarding & 10 & 2.44 & 3 & 0.73 & 3 & 0.73 & 3 & 0.73 & 1 & 0.24 \\
         & Serialization Fault & 9 & 2.20 & 1 & 0.24 & 2 & 0.49 & 5 & 1.22 & 1 & 0.24 \\
         & Concurrency Fault & 6 & 1.47 & 0 & 0.00 & 1 & 0.24 & 5 & 1.22 & 0 & 0.00 \\
        & Subtotal & 212 & 51.83 & 70 & 17.11 & 40 & 9.78 & 64 & 15.65 & 38 & 9.29 \\
        Agent-Specific Fault & Model-Related Fault & 48 & 11.74 & 11 & 2.69 & 1 & 0.24 & 28 & 6.85 & 8 & 1.96 \\
         & Cognitive Context Mismanagement & 24 & 5.87 & 2 & 0.49 & 6 & 1.47 & 7 & 1.71 & 9 & 2.20 \\
         & Orchestration Fault & 16 & 3.91 & 2 & 0.49 & 5 & 1.22 & 8 & 1.96 & 1 & 0.24 \\
        & Subtotal & 88 & 21.52 & 15 & 3.67 & 12 & 2.93 & 43 & 10.51 & 18 & 4.40 \\
        Documentation Fault & Documentation Fault & 51 & 12.47 & 8 & 1.96 & 8 & 1.96 & 16 & 3.91 & 19 & 4.65 \\
        Dependency/Environment Fault & Dependency Fault & 35 & 8.56 & 10 & 2.44 & 9 & 2.20 & 12 & 2.93 & 4 & 0.98 \\
         & Environment Incompatibility & 12 & 2.93 & 0 & 0.00 & 5 & 1.22 & 5 & 1.22 & 2 & 0.49 \\
        & Subtotal & 47 & 11.49 & 10 & 2.44 & 14 & 3.42 & 17 & 4.16 & 6 & 1.47 \\
        User Misuse & User Misuse & 11 & 2.69 & 2 & 0.49 & 3 & 0.73 & 6 & 1.47 & 0 & 0.00 \\
    \end{tblr}
\end{table*}

Table~\ref{tab:root-cause} summarizes root-cause distributions across frameworks, including raw counts (\#) and proportions (\%). 
\textbf{General Programming Fault} accounts for the largest share (51.83\%), with Missing Case Handling (12.47\%) as the most frequent subcategory. Its consistent presence across frameworks suggests a systemic lack of defensive programming practices rather than framework-specific design issues, making it largely amenable to conventional debugging techniques.
\textbf{Agent-Specific Fault} accounts for 21.52\% of all bugs, comprising Model-Related Fault (11.74\%), Cognitive Context Mismanagement (5.87\%), and Orchestration Fault (3.91\%).These subcategories reflect challenges unique to agentic frameworks, spanning model integration, context management, and multi-agent coordination. \emph{Model-Related Fault are nearly as prevalent as Missing Case Handling.}
Notably, \autogen accounts for the majority of Model-Related Fault (6.85\% out of 11.74\%), likely due to its support for heterogeneous model backends, which increases compatibility risks.
\textbf{Documentation Fault} (12.47\%) ranks third, primarily caused by incorrect or outdated example code, reflecting challenges in keeping documentation synchronized with rapidly evolving agentic APIs. 

Compared to the baseline~\cite{xue2025llmbugstudy}, which identifies Code Logic Issues (38\%) and API Misuse (25\%) as dominant in early LLM pipelines, our analysis reveals a structural shift in agentic frameworks. We decouple \textbf{Model-Related Fault} from generic API Misuse to capture adaptation gaps between framework logic and heterogeneous model protocols. Meanwhile, coordination-related issues previously subsumed under code logic are distinguished into \textbf{Cognitive Context Mismanagement} and \textbf{Orchestration Fault}, both absent from the baseline taxonomy.

\begin{findingbox}
\textbf{Finding 2:}
While General Programming Fault dominates (52\%), Agent-Specific Fault (22\%) represents root cause categories absent from the baseline study, comprising Model-Related Fault, Cognitive Context Mismanagement, and Orchestration Fault.
\end{findingbox}

\section{RQ3: Bug-prone Components}
\label{sec:component}

\begin{table}[t] 
    \centering
    \footnotesize
    \caption{Bug distribution by components and frameworks.}
    \label{tab:component}
    \begin{tblr}{
            width = \columnwidth,
            colspec = {l r r r r r r r r r r r r}, 
            cells = {m},
            colsep = 1.5pt, 
            rowsep = 1.5pt, 
            cell{1}{2,4,6,8,10,12} = {c=2}{m}, 
            cell{1}{1} = {r=2}{m},
            hline{1,2,3,Y,Z} = {solid}, 
            vline{2,6,8,10,12} = {solid}, 
            vline{4,X} = {0.8pt, solid},
            row{even} = {bg=gray!10}, 
            row{1,2} = {c, font=\bfseries, bg=gray!30},
            row{Z} = {bg=gray!30},
            cell{Z}{1} = {font=\bfseries},
            cell{3,4}{2,3} = {font=\bfseries},
            cell{3,6}{4,5} = {font=\bfseries},
            cell{4,7}{6,7} = {font=\bfseries},
            cell{3,4}{8,9} = {font=\bfseries},
            cell{5,8}{10,11} = {font=\bfseries},
            cell{4,5}{Y,Z} = {font=\bfseries},
        }
        Component & Total & & LC/LG & & \crewai & & \autogen & & Smol & & Tested & \\
        & \# & \% & \# & \% & \# & \% & \# & \% & \# & \% & \# & \% \\
        Intelligence & 101 & 24.69 & 38 & 9.29 & 6 & 1.47 & 42 & 10.27 & 15 & 3.67 & 47 & 46.53 \\
        Orchestration & 82 & 20.05 & 7 & 1.71 & 23 & 5.62 & 36 & 8.80 & 16 & 3.91 & 54 & 65.85 \\
        Action & 62 & 15.16 & 7 & 1.71 & 9 & 2.20 & 22 & 5.38 & 24 & 5.87 & 38 & 61.29 \\
        Knowledge & 56 & 13.69 & 30 & 7.33 & 13 & 3.18 & 10 & 2.44 & 3 & 0.73 & 25 & 44.64 \\
        Infrastructure & 54 & 13.20 & 14 & 3.42 & 16 & 3.91 & 19 & 4.65 & 5 & 1.22 & 17 & 31.48 \\
        Documentation & 54 & 13.20 & 9 & 2.20 & 10 & 2.44 & 17 & 4.16 & 18 & 4.40 & 0 & 0.00 \\
        Total & 409 & 100.00 & 105 & 25.70 & 77 & 18.80 & 146 & 35.70 & 81 & 19.80 & 181 & 44.30 \\
    \end{tblr}
\end{table}

Table~\ref{tab:component} shows the bug distribution across framework components. While our analysis primarily focuses on functional components, we also include ``Documentation'' to provide a comprehensive view of all reported issues. We included a \emph{Tested} column to measure the percentage of bug fixes accompanied by test cases. 
Bugs are predominantly found in core functional components, with Intelligence accounting for the largest share (25\%), followed by Orchestration (20\%) and Action (15\%). These results indicate that model adaptation and coordination logic are the primary sources of fault. 
This distribution further varies across frameworks according to their design priorities. For instance, \smolagents skews toward Action (6\%) due to its action-in-code design, and LC/LG shows a notably higher proportion in  Knowledge (7\%) due to its RAG emphasis.

While the baseline study~\cite{xue2025llmbugstudy} identified Data Processing (40\%) and Core Schema (24\%) as the key bug-prone components in early pipelines, our results show the reliability bottleneck has shifted to Intelligence and Orchestration in modern agentic frameworks, reflecting the increasing complexity of multi-agent coordination. 
Our five-layer abstraction pinpoints bugs in model adaptation and coordination logic that are collapsed together in prior study.



Our analysis also reveals a significant misalignment between empirical risk and testing effort during the bug-fixing process. While Intelligence is the most bug-prone, it receives a disproportionately low test inclusion rate (47\%), whereas Orchestration and Action see much higher rates (66\% and 61\%).



\begin{findingbox}
\textbf{Finding 3:} 
Intelligence (25\%) and Orchestration (20\%) are the most bug-prone layers overall, though the dominant components differ across frameworks. Developers include tests more often for Orchestration than Intelligence (66\% vs.\ 47\%), leaving the most bug-prone layer undervalidated.

\end{findingbox}

\section{RQ4: Commonality \& Association}
\label{sec:common_assoc}

\subsection{Cross-framework Commonality}



All studied frameworks show high cross-framework commonality in symptoms (0.76--0.86) and root causes (0.80--0.88). Component-level similarity (0.62--0.81) is lower, largely due to \langchain's disproportionate concentration of bugs in the Knowledge components (Section~\ref{sec:component}). Nevertheless, the lower bound of 0.62 still indicates substantial commonality overall. These patterns suggest a layered evaluation framework: universal black-box oracles for common symptoms, complemented by white-box testing targeting framework-specific bug-prone components to improve diagnosis and coverage.

\begin{figure}[!htbp]
    \centering
    \begin{subfigure}[b]{0.32\columnwidth}
        \centering
        \includegraphics[width=\linewidth]{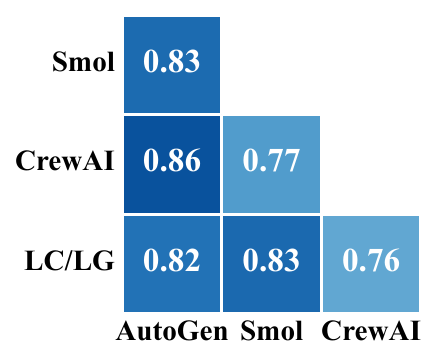}
        \caption{Symptoms}
        \label{fig:common_symp}
    \end{subfigure}
    \hfill
    \begin{subfigure}[b]{0.32\columnwidth}
        \centering
        \includegraphics[width=\linewidth]{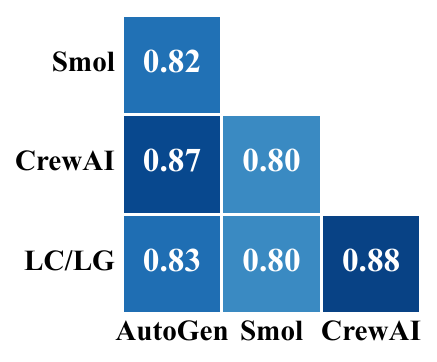}
        \caption{Root Causes}
        \label{fig:common_rc}
    \end{subfigure}
    \hfill
    \begin{subfigure}[b]{0.32\columnwidth}
        \centering
        \includegraphics[width=\linewidth]{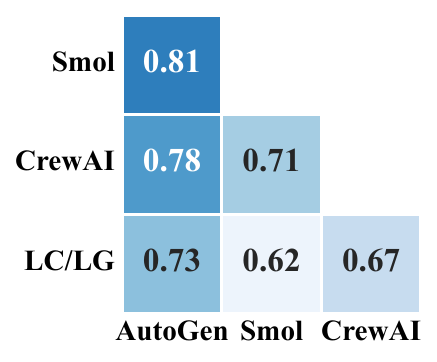}
        \caption{Components}
        \label{fig:common_comp}
    \end{subfigure}

    \caption{Jensen-Shannon similarity between frameworks.}
    \label{fig:commonality}
\end{figure}




\subsection{Inter-dimensional Relationships}
\begin{table}[t]
    \centering
    \small
    \begin{talltblr}[
        theme = mockcaption,
        caption = {\textbf{Significant co-occurring category pairs.}},
        label = {tab:relation},
        remark{Abbreviations} = {
            \textit{Init. Failure}, \textit{Unexp. State} and \textit{Unexp. Sequence} denote Initialization Failure, Unexpected Intermediate State and Unexpected Execution Sequence, respectively.
        },
    ]{
        width = \columnwidth,
        colspec = {l l l r },
        cells = {m}, 
        colsep = 3pt, 
        rowsep = 1.5pt, 
        cell{2}{1} = {r=2}{m},
        cell{4}{1} = {r=5}{m},
        cell{9}{1} = {r=4}{m},
        cell{4,7}{2} = {r=2}{m},
        cell{W,Y}{3} = {r=2}{m},
        hlines,
        vline{2,3,4} = {solid}, 
        column{1} = {bg=white},   
        row{1} = {font=\bfseries, bg=gray!30}, 
    }

Relationship & Category A & Category B  & $N$  \\

{Root Cause $\times$\\ Symptom} & Agent-specific Fault & Unexp. State & 14 \\
 & Dependency/Environment Fault & Init.  Failure & 30 \\
{Root Cause $\times$\\ Component} & Agent-specific Fault & Intelligence & 44 \\
 &  & Orchestration & 32 \\
 & Dependency/Environment Fault & Infrastructure & 21 \\
 & General Programming Fault & Action & 52 \\
 &  & Knowledge & 44 \\
{Symptom $\times$\\ Component} & Init.  Failure & Infrastructure & 20 \\
 & Incomplete/Incorrect Trace & Infrastructure & 6 \\
 & Unexp. Sequence & Orchestration & 12 \\
 & Unexp. State & Orchestration & 14 \\
    \end{talltblr}
\end{table}

Chi-squared tests confirm significant associations ($p < 0.001$) across all analyzed dimension pairs, indicating that bug characteristics are interdependent. Strongest associations involve Root Cause with Symptom ($\chi^2 = 241.19, V = 0.384$) and Component ($\chi^2 = 472.56, V = 0.537$), while the association between Symptom and Component is also substantial ($\chi^2 = 171.16, V = 0.289$). 

Table~\ref{tab:relation} lists significant category pairs from post-hoc analysis ($N$ denotes frequency). We omit pairs involving Others symptom or documentation-only issues. 
These associations reveal two bug patterns newly observed in agentic frameworks. First, faults in Orchestration components frequently manifest as Unexpected Intermediate State and Unexpected Execution Sequence, indicating that frameworks often fail to maintain state and sequence integrity across complex topologies. For instance, in \crewai-1463~\cite{crew1463multistartflow}, the workflow failed to handle multi-start configurations, leading to untracked execution paths~\cite{crew1463multistartflow}.
Second, the association between Action/Knowledge components and General Programming Fault highlights a lack of defensive logic in deterministic components when handling edge cases triggered by LLM interactions. 
These relationships enable the design of symptom-based test oracles and targeted defensive mechanisms to mitigate the specific root causes. 

\begin{findingbox}
\textbf{Finding 4:}
Symptoms, root causes, and components show substantial cross-framework commonality (0.62--0.88) and significant inter-dimensional associations. Category-level analysis reveals two core bug patterns:
(1) Orchestration components frequently compromise state and sequence integrity, and (2) Action/Knowledge components often lack defensive logic against LLM-driven interactions.

\end{findingbox}

\section{RQ5: Triggering Scenarios}
\label{sec:triggering}


\begin{figure}[t]
\centering
    \begin{subfigure}[t]{0.42\textwidth}
    \centering
        \begin{pythoncode}
# Element Config. Labels: "Model:Backend" and "Model:ModelID"
client = !\hladd{OpenAIClient}!(!\hladd{model}!="gemini-1.5-flash")
# Input Characteristic Label: "InputMessages" 
messages = [
    !\hladd{SystemMessage}!(content="Say anything Start with 'FOO'"),
    !\hladd{SystemMessage}!(content="Say anything End with 'BAR'"),
    UserMessage(content="Just say '.'", source="user")]
result = await client.create(messages=messages)
        \end{pythoncode}
        \caption{\autogen-6116~\cite{ag6116multisysprompt} \label{fig:modelbackendid}}
    \end{subfigure}
    \hfill
    \begin{subfigure}[t]{0.42\textwidth}
    \centering
        \begin{pythoncode}
@tool # Element Configuration Label: "Tool:Signature"
def calculator(!\hladd{args: Dict[str, str]}!):
    pass

agent = create_react_agent(model, tools=[calculator])
executor = AgentExecutor(agent, tools=[calculator])
# Operation Label: TaskIntent:CalculationTask
result = executor.invoke({"input": "10+25=?"})
# Input Characteristic Label: "LLMResponse:ToolCallSchema"
# Non-standard JSON formatting: "{'a':'10', 'b':'25', 'operator':'+'}"
        \end{pythoncode}
        \caption{\langchain-30910~\cite{lc30910toolsig} \label{fig:toolsig}}
    \end{subfigure}
    \caption{Simplified triggering scenarios with labels.}
\end{figure}

{
\newcommand{\mygap}{\vspace{0.5em}}

This section analyzes frequent bug-triggering factors among 273 bugs across three dimensions: element configurations, input characteristics, and operations. Factors are not mutually exclusive, as a single bug may involve multiple triggers. Detailed definitions of these factors are provided in Appendix~\ref{sec:appendix-factors}.

\mygap 
\noindent \textbf{\textit{Element Configurations}(251, 92\%).}
Element configurations are the most pervasive dimension, involved in 92\% of bug-triggering scenarios. Model (88, 35\%) settings are the most frequent trigger, with Backend (55, 22\%) and ModelID (39, 16\%) governing communication protocols and LLM identity, respectively. Tool (64, 25\%) definitions shape how agents interact with external services, where BuiltinType (18, 7\%) and Signature (16, 6\%) misconfigurations directly corrupt tool-calling prompts. Beyond these, Agent (38, 15\%), Team (30, 12\%), CodeExecutor (17, 7\%), and Store (15, 6\%) cover coordination, execution, and persistence concerns, respectively.
Notably, pattern mining identifies a high-risk combination where bugs emerge exclusively from specific Backend and ModelID co-configurations (8\%), exemplified by \autogen-6116 triggered by the \texttt{OpenAIClient} and ``gemini-1.5-flash'' pairing (Figure ~\ref{fig:modelbackendid}).

\mygap 
\noindent \textbf{\textit{Input Characteristics} (56, 21\%).}
TaskIntent (19, 34\%) and LLM\-Response (12, 21\%) are the two dominant triggers. Complex task intents---such as MultiToolCallTask (5, 9\%) and MultimodalTask (5, 9\%)---frequently expose flaws in orchestration logic; for instance, in \smolagents-1481, a prompt to ``launch two web search agents in parallel'' exposed greedy behavior in the manager agent, causing incomplete answers~\cite{smol1481multicall}. LLMResponse-related triggers arise when frameworks fail to handle atypical model outputs, namely ToolCallSchema (6, 11\%) and RegularSchema (5, 9\%) violations; as shown in Figure~\ref{fig:toolsig}, a non-standard JSON string in the arguments field causes parsing failures. InputMessages (11, 20\%) captures structural or formatting anomalies in message history, frequently arising from model-specific constraints---such as Anthropic models rejecting multiple system messages (Figure ~\ref{fig:modelbackendid})---that frameworks fail to uniformly enforce. Code (8, 14\%) covers executable code inputs whose specific patterns expose implementation limitations of code execution components; for example, in \smolagents-839, chained assignments crashed the \texttt{LocalPythonExecutor}.

\mygap 
\noindent \textbf{\textit{Operations} (101, 37\%).} 
CallTool (24, 24\%) is the most frequent operation trigger, reflecting that the interface between agents and external services is a high-risk boundary. ExportComponent (16, 16\%) and LoadComponent (7, 7\%) highlight serialization and deserialization as risk-prone operations, where corruption in the dump or load process can compromise framework state. ExecCode (11, 11\%) involves dynamic execution of generated code within framework-managed sandboxes. ConsecutiveExec (8, 8\%) exposes residual state risks from repeated execution of the same operation—as shown in \autogen-6746~\cite{ag6746consecexec}, residual state from previous runs corrupts the context of subsequent calls. ProcessDoc (7, 7\%) covers utility operations for handling external documents.

Cross-dimension pattern mining yields no combinations meeting the support threshold ($\ge$ 5\%), suggesting that triggering conditions across dimensions tend to occur independently.
}

\begin{findingbox}
\textbf{Finding 5:} 
Element configurations are the most pervasive trigger dimension (92\%), with pattern mining identifying specific Backend and ModelID co-configurations as a high-risk combination (8\%).
Complex TaskIntent (34\%) and atypical LLMResponse (21\%) dominate input-related triggers, with InputMessages bugs frequently tied to model-specific message formatting constraints.
Among operations, CallTool (24\%) is the most frequent trigger, while serialization and consecutive execution further highlight state management as a recurring risk.
\end{findingbox}

\section{RQ6: Transferability}
\label{sec:transfer}

We found 16 of 35 source bugs (46\%) involving shared agentic interaction surfaces are transferable, yielding 29 issues across all studied frameworks: 4 Duplicates, 11 Adapted, and 14 Variants. Duplicates and Adapted cases account for over half of transfers, indicating many functional counterparts across frameworks. Variants further show that core triggering patterns generalize across components to uncover latent faults, supporting cross-framework bug transfer.

We submitted 11 bug reports, all acknowledged by maintainers or community contributors. One was fixed, six received community acknowledgment via reproductions or fixes, and four were recognized by maintainers as design boundaries or addressed with alternative solutions. These varied resolutions reflect genuine design trade-offs rather than dismissals---for example, we adapted a serialization bug from \autogen-6855~\cite{ag6855apimiuse} to \langchain-34925~\cite{lg34925adapted},  where AI messages in chat history lost type-specific fields like tool call details; maintainers acknowledged the issue but suggested migrating to newer memory abstractions in \langgraph rather than fixing it directly. Such resolutions reflect genuine design trade-offs rather than dismissals, showing that our cross-framework transfers effectively uncover community-recognized faults across diverse implementations.

\begin{findingbox}
\textbf{Finding 6:} Nearly half of sampled bugs involving shared agentic interaction surfaces are transferable across frameworks as adapted functional counterparts or generalized variants, highlighting that existing bug-triggering patterns provide effective templates for cross-framework validation.
\end{findingbox}

\section{Discussion}

\noindent \textbf{From Architectural Evolution to Interaction-Level Reliability.}
Modern agentic frameworks have evolved from passive libraries to active orchestrators governing state, control flow, and agent coordination, expanding the correctness concerns from individual function outputs to include multi-step interactions among agents, models, and external services. Consistent with Findings 2 and 3, the most bug-prone layers and root causes are concentrated around model integration and workflow coordination, indicating that reliability challenges increasingly arise at interaction boundaries rather than within isolated implementations. Model-Related Fault occurs nearly as frequently as Missing Case Handling, while the growing prevalence of Initialization Failure (Section~\ref{sec:symptom-analysis}) further highlights increasing dependence on external services and runtime ecosystems. Accordingly, our Agent-Specific Fault categories provide structured diagnostic paths for localizing interaction-related faults.

\noindent \textbf{Validation Gaps: Testing Bias and Oracle Limits.}
Despite being the most bug-prone layer, Intelligence exhibits a notable mismatch between bug density and test inclusion rate (Finding 3). This aligns with prior observations that developers preferentially test deterministic components~\cite{hasan2025empirical}, and our data extend this by showing the bias persists during bug-fixing itself---suggesting that testing toolchains remain inadequately equipped to mock or simulate complex external model dependencies. The newly identified symptom categories further expose structural oracle limitations (Finding 1): Unexpected Execution Sequence and User Configuration Ignored indicate that conventional fuzzing and assertion-based testing are insufficient to validate execution order deviations or configuration enforcement, respectively.

\noindent \textbf{Cross-Framework Benchmarking and Transferability.}
The substantial cross-framework commonality in symptoms and root causes (Finding 4), together with the transferability of nearly half of sampled bugs and developer engagement with all submitted reports (Finding 6), provides strong evidence for reusable validation artifacts across agentic frameworks. This evidence supports a layered evaluation framework in which shared benchmarks capture common interaction-level correctness concerns, while framework-specific extensions address distinctive architectural mechanisms. Such a framework could enable community-maintained benchmark suites for systematic cross-framework validation.

\noindent \textbf{Design, Testing, and Research Implications.}
Collectively, our findings suggest that reliability in agentic frameworks should be treated as an interaction-contract problem. For designers, this calls for defining explicit interaction assumptions through typed interfaces and compatibility contracts, while making execution-state transitions and termination criteria explicit at the interaction boundary. For testers, the observed validation gaps and configuration-dependent faults motivate interaction-aware testing strategies that systematically exercise execution sequences, configuration spaces, and atypical LLM behaviors. For researchers, the persistent oracle challenges and cross-framework bug transferability highlight opportunities for automated oracle construction, configuration-aware testing techniques, and community-driven validation benchmarks.

\section{Threats to Validity}
We identify three threats to validity. 
First, subjective bias in manual annotation was mitigated through multi-author coding, achieving high inter-annotator agreement (Cohen's $\kappa$ > 0.89).
Second, to ensure generalizability, we analyzed \datacount fixed bugs across five frameworks representing diverse architectural paradigms, covering a broad ecosystem. 
While LangChain issues were randomly downsampled for cross-framework balance, our goal is taxonomy construction rather than prevalence estimation; the consistent patterns observed across frameworks support the robustness of the taxonomy.
Finally, to prevent misclassification of framework-level bugs, we applied a three-tier definition (task, application, and framework bugs) and analyzed only confirmed fixes from official repositories. 
\section{Related Works}
\noindent \textbf{Task- and Application-level Failures.}
Recent studies examine agent failure modes at the task level, focusing on inter-agent misalignment~\cite{pan2025whymast} or planning and execution errors~\cite{lu2025exploretask}. Others explore the application layer, identifying bugs in prompt engineering or end-user logic~\cite{islam2026whenlabeling}. While these studies offer valuable insights into agent behavior, they primarily analyze model-driven symptoms or application-specific faults. Our study targets the framework layer, investigating internal source code and architectural defects of the orchestration engines that underpin these systems.

\noindent \textbf{Framework-level Bug Characterization.}
While Xue et al. characterize bugs in early LLM pipelines~\cite{xue2025llmbugstudy}, our work addresses the complexities of modern multi-agent systems through a more granular five-layer architectural model. We move beyond general programming errors to identify specialized agent-specific symptoms and root causes (\eg Cognitive Context Mismanagement). We further evaluate bug-triggering scenarios and their cross-framework transferability, offering a predictive perspective on structural vulnerabilities that are absent in prior literature.

\section{Conclusions}
This paper presents an empirical study of \datacount fixed bugs across five modern agentic frameworks, analyzing their symptoms, root causes, bug-prone components, triggering conditions, and cross-framework transferability under a five-layer architectural model. 
The results indicate a paradigm mismatch in reliability assessment: correctness in agentic systems depends on interaction contracts across agents, models, and external services, whereas prevailing testing techniques (\eg fuzzing and assertion-based oracles) are designed for deterministic, function-level behaviors and therefore fail to capture interaction-driven failures. 
Cross-framework commonality in symptoms and root causes (0.62--0.88), together with substantial bug transferability, further shows that these issues stem from shared architectural interaction patterns rather than framework-specific defects, motivating unified testing infrastructure. 
A key limitation is that our dataset, derived from resolved GitHub issues, excludes silent and unreported failures, which may affect distributional generality. 
We identify three directions for future work: oracles for execution-order and configuration constraints, mock infrastructure for heterogeneous LLM behaviors, and cross-framework benchmarks based on transferable triggering patterns.

\noindent \textbf{Data Availability Statement.} Data, code, and the annotation cookbook are available at~\cite{dr_data_code}.

\begin{acks}
This work was supported by the
\grantsponsor{NSERC}{Natural Sciences and Engineering Research Council of Canada}{https://doi.org/10.13039/501100000038}
under Discovery Grant No.~\grantnum{NSERC}{RGPIN-2024-04301}.
\end{acks}

\appendix

\section{Detailed Definitions of Triggering Factors}
\label{sec:appendix-factors}

\subsection{Element Configurations (251, 92\%).}
\noindent \textbf{Model (88, 35\%):} Model clients used to interface with LLMs. 
\begin{tightitemize}
\item \textbf{Backend (55, 22\%):} The internal client class  managing communication protocols (\eg \texttt{OpenAIClient} in Figure~\ref{fig:modelbackendid}).

\item \textbf{ModelID (39, 16\%):} The specific LLM identifier executing the reasoning tasks (\eg \texttt{gemini-1.5-flash}).
\end{tightitemize}

\noindent \textbf{Tool (64, 25\%):} This enables agents to take actions.
\begin{tightitemize}
\item \textbf{BuiltinType (18, 7\%):} Pre-implemented tool classes providing ready-to-use actions (\eg \texttt{AzureAISearchTool} in \autogen).
\item \textbf{Signature (16, 6\%):} The parameter types and return schema used by the framework to structure tool-calling prompts.
\end{tightitemize}

\noindent\textbf{Agent (38, 15\%):} Autonomous entities orchestrating reasoning and tool 
usage, characterized by role and planning logic. 

\noindent\textbf{Team (30, 12\%):} High-level entities coordinating multi-agent 
collaboration via hierarchical or flat organizational structures.



\noindent \textbf{CodeExecutor (17, 7\%):} It manages the execution environment for generated code snippets. For example, \texttt{ExecutorType} includes runtime backend for execution (e.g., local or remote executors). 

\noindent \textbf{Store (15, 6\%):} Data persistence and indexing, with a frequent attribute \texttt{Backend} representing the underlying database or file system.

\subsection{Input Characteristics (56, 21\%).}
Characteristics of user requests or model outputs that trigger faults.

\noindent \textbf{TaskIntent (19, 34\%):} Abstract patterns representing the functional goal of a request. 
\begin{tightitemize}
\item \textbf{MultiToolCallTask  (5, 9\%):} Requests for orchestrating multiple tool or agent calls. In \smolagents-1481, a prompt to ``launch two web search agents in parallel''  exposed greedy behavior in the manager agent, causing incomplete answers~\cite{smol1481multicall}. 

\item \textbf{MultimodalTask  (5, 9\%):} Tasks involving non-textual data (\eg images), provided either as input or generated as tool outputs.
\end{tightitemize}

\noindent \textbf{LLMResponse (12, 21\%):} Patterns in model outputs that deviate from framework expectations. 
\begin{tightitemize}
\item \textbf{ToolCallSchema (6, 11\%):} Atypical or malformed tool call messages that bypass parsers. In Figure~\ref{fig:toolsig} (Line 10), the LLM generates a non-standard JSON string for the \texttt{arguments} field, leading to parsing failures and subsequent invocation errors.
\item \textbf{RegularSchema  (5, 9\%):} Flaws or non-determinism in standard text responses that break downstream logic (\eg non-deterministic text formats in \langchain-8716~\cite{lc8716ndformat}). 
\end{tightitemize}

\noindent \textbf{InputMessages (11, 20\%):} Structural or formatting features of the message history, such as multiple system messages in Figure~\ref{fig:modelbackendid}.

\noindent \textbf{Code (8, 14\%):} Specific patterns in code snippets that expose implementation limitations of executors. In \smolagents-839, chained assignments crashed the \texttt{LocalPythonExecutor}. 

\subsection{Operations (101, 37\%).}
Bug-triggering runtime actions.

\noindent \textbf{CallTool (24, 24\%):} Framework invocations of external tools. 

\noindent \textbf{ExportComponent (16, 16\%), LoadComponent (7, 7\%):} Operations for serializing and restoring framework entities, such as state corruption in the \texttt{dump} or \texttt{load\_component}~\cite{ag6336serial}. 

\noindent \textbf{ExecCode (11, 11\%):} Dynamic execution of generated code within framework-managed sandboxes.

\noindent \textbf{ConsecutiveExec (8, 8\%):} Repeated execution of the same task or instance. In \autogen-6746~\cite{ag6746consecexec} residual state from previous runs corrupts the context, causing subsequent calls to fail.

\noindent \textbf{ProcessDoc (7, 7\%):} Utility operations for handling external documents, such as batch indexing (\eg \langchain-32272~\cite{lc32272procdocs}).


\bibliographystyle{ACM-Reference-Format}
\bibliography{refs}


\end{document}